\documentclass[twocolumn,10pt,prl,letterpaper,groupedaddress]{revtex4}

\usepackage{times,amsmath,amsfonts,amssymb,latexsym,textcomp}
\usepackage{color}
\usepackage{graphicx}
\usepackage{multirow}
\usepackage{bbm}
\usepackage{tabularx}
\usepackage{amsthm}
\usepackage{dsfont}
\usepackage{placeins}
\usepackage[lofdepth,lotdepth]{subfig}
\usepackage[T1]{fontenc}
\usepackage[latin9]{inputenc}
\usepackage[english]{babel}
\usepackage{float}
\usepackage{array}
\usepackage{caption}
\usepackage{braket}
\usepackage{MnSymbol}
\usepackage{tikz}
\usepackage[electronic]{ifsym}
\usepackage{booktabs}
\usepackage{subfig}

\renewcommand{\phi}{\varphi}

\renewcommand{\epsilon}{\varepsilon}

\begin{document}
 \title{The Minimum Complexity of Kochen-Specker Sets Does Not Scale with Dimension}
\author{Mordecai Waegell$^1$ and P. K. Aravind$^2$}
\affiliation{$^1$~Institute for Quantum Studies, Chapman University, Orange, CA, US \\
$^2$~Worcester Polytechnic Institute, Worcester, MA, US}

\begin{abstract}
A Kochen-Specker (KS) set is a specific set of projectors and measurement contexts that prove the Bell-Kochen-Specker contextuality theorem.  The simplest known KS sets in Hilbert space dimensions $d=3,4,5,6,8$ are reproduced, and several methods by which a new KS set can be constructed using one or more known KS sets in lower dimensions are reviewed and improved.  These KS sets and improved methods enable the construction of explicitly {\it critical} new KS sets in all dimensions, where critical refers to the irreducibility of the set of contexts.  The simplest known critical KS sets are derived in all even dimensions $d\geq10$ with at most 9 contexts and 30 projectors, and in all odd dimensions $d\geq 7$ with at most 13 contexts and 39 projectors.  These results show that neither the number of contexts nor the number of projectors in a minimal KS set scales with dimension $d$.
\end{abstract}
\maketitle


\textbf{Introduction}:--- Quantum contextuality, as first conceived by Bell, Kochen, and Specker \cite{Bell2, KS}, is an important subject in the study of the foundations of quantum mechanics, and has also been receiving recent attention because of its connection to quantum advantage in quantum information processing \cite{galvao2005discrete, spekkens2008negativity, howard2014contextuality, bermejo2016contextuality}.

Quantum contextuality refers to the inability of a noncontextual hidden variable theory (NCHVT) to make exact predictions for the outcomes of all projective measurements that also reproduce the predictions of quantum mechanics.  A set of mutually orthogonal projectors may be simultaneously measured, and we call any complete set of mutually orthogonal projectors a {\it context}, where completeness indicates that the set of projectors spans the system Hilbert space.  A context defines a specific measurement procedure on the system, with each projector corresponding to a particular yes/no test that will be performed as part of the measurement.  Noncontextuality is the assumption from classical physics that the predicted value of a given projector must be independent of what context it is measured in (i.e., independent of which other commuting projectors may be measured simultaneously).

The Kochen-Specker (KS) theorem proves that NCHVTs cannot be consistent with quantum mechanics by exhibiting a discrete set of projectors for which no  noncontextual value assignment of 0 or 1 to all of the projectors is possible without violating the quantum rules that exactly one projector in every complete context comprised by the set must be assigned a 1, and no two orthogonal projectors can both be assigned a 1.

Specific sets of projectors with this property are called {\it KS sets} and the first, containing 117 projectors in dimension $d=3$, was given by Kochen and Specker \cite{KS}.  Since then, many more KS sets have been found in all dimensions $d\geq 3$ \cite{penrose2000bell, zimba1993bell, peres1993quantum, Peres24, Mermin_SquareStar, kernaghan1994kochen, Cabello18_9, cabello2005recursive,matsuno2007construction, WA_600cell_1, WA_600cell_2, MP_600cell, pavivcic2010new,yu2011minimal, WA_24Rays, WA_60Rays, WA_3qubits, WA_Nqubits, waegell2014parity,lisonvek2014kochen, lisonek2014generalized, waegell2015parity}, some simpler, and some more exotic.  The set with the fewest projectors has just 18, which comprise 9 complete contexts in $d=4$ \cite{Cabello18_9}, while the set with the fewest complete contexts has just 7, comprised of 21 projectors in $d=6$ \cite{lisonvek2014kochen}, and there are good reasons to believe that these are the simplest possible cases.

Because each projector in both of these KS sets belongs to two contexts, these sets also have a particular property, called {\it parity}, which makes the impossibility of NCHVT 0/1 assignments easy to see.  A KS parity set is one in which the $R$ projectors comprise an odd number $B$ of complete contexts in such a way that each projector belongs to an even number of the contexts.  Then a complete 0/1 assignment to all contexts would need to have an odd number $B$ of 1s --- one for each context, in order to obey the quantum prediction, but an even number in order to obey noncontextuality, making such an assignment impossible.  The impossibility of noncontextual 0/1 assignments to KS sets without parity is not generally obvious and is verified through exhaustive computational checks.

The simplest KS sets are also critical, where a set is {\it critical} if removing any one context from it allows a noncontextual 0/1 assignment to the set of projectors such that exactly one projector is assigned the value 1 in each of the remaining contexts.  Note that this definition puts the focus on minimal sets of complete measurement contexts rather than minimal sets of projectors.  All of the KS sets that we discuss in this letter are critical KS sets, unless otherwise noted, and we will often drop the extra descriptor for brevity.  Criticality is not generally obvious and must be verified through an exhaustive computational check.


The remainder of this letter is organized as follows:  In the next section we review the smallest known KS sets in dimensions $d=3,4,6,8$ and discuss their properties.  In the following section we review and improve methods introduced by Penrose and Zimba \cite{zimba1993bell}, Cabello, Estebaranz, and Garc{\'\i}a-Alcaine\cite{cabello2005recursive}, and Matsuno \cite{matsuno2007construction} for constructing new KS sets from one or more known KS sets in lower dimensions.   We then apply these methods to the known KS sets in $d=4,6,8$ to generate and catalog critical KS sets in all dimensions $d \geq 5$, and conclude with a few remarks about the implications of this work.

\begin{figure}
\centering
\subfloat[][]{
\label{18_9}
\begin{tabular}{c|cccc}
1 &  $1$ &  $0$ &  $0$ &  $0$ \\
2 &  $0$ &  $1$ &  $0$ &  $0$ \\
3 &  $0$ &  $0$ &  $1$ &  $0$ \\
4 &  $1$ &  $1$ &  $1$ &  $1$ \\
5 &  $1$ &  $1$ &  $1$ &  $\bar{1}$ \\
6 &  $1$ &  $\bar{1}$ &  $1$ &  $1$ \\
7 &  $1$ &  $\bar{1}$ &  $1$ &  $\bar{1}$ \\
8 &  $1$ &  $\bar{1}$ &  $\bar{1}$ &  $1$ \\
9 &  $1$ &  $\bar{1}$ &  $\bar{1}$ &  $\bar{1}$ \\
10 &  $1$ &  $1$ &  $0$ &  $0$ \\
11 &  $1$ &  $0$ &  $0$ &  $1$ \\
12 &  $1$ &  $0$ &  $0$ &  $\bar{1}$ \\
13 &  $1$ &  $0$ &  $\bar{1}$ &  $0$ \\
14 &  $0$ &  $1$ &  $0$ &  $1$ \\
15 &  $0$ &  $1$ &  $0$ &  $\bar{1}$ \\
16 &  $0$ &  $1$ &  $\bar{1}$ &  $0$ \\
17 &  $0$ &  $0$ &  $1$ &  $1$ \\
18 &  $0$ &  $0$ &  $1$ &  $\bar{1}$ \\
\end{tabular}}
\qquad
     \subfloat[][]{\label{21_7}
\begin{tabular}{c|cccccc}
1 &  $1$ &  $1$ &  $1$ &  $1$ &  $1$ &  $1$ \\
2 &  $1$ &  $1$ &  $\bar{\omega}$ &  $\omega$ &  $\omega$ &  $\bar{\omega}$ \\
3 &  $1$ &  $\bar{\omega}$ &  $1$ &  $\omega$ &  $\bar{\omega}$ &  $\omega$ \\
4 &  $1$ &  $\bar{\omega}$ &  $\omega$ &  $\bar{\omega}$ &  $\omega$ &  $1$ \\
5 &  $1$ &  $\omega$ &  $\omega$ &  $1$ &  $\bar{\omega}$ &  $\bar{\omega}$ \\
6 &  $1$ &  $\omega$ &  $\bar{\omega}$ &  $\bar{\omega}$ &  $1$ &  $\omega$ \\
7 &  $1$ &  $1$ &  $\omega$ &  $\bar{\omega}$ &  $\bar{\omega}$ &  $\omega$ \\
8 &  $1$ &  $\omega$ &  $1$ &  $\bar{\omega}$ &  $\omega$ &  $\bar{\omega}$ \\
9 &  $1$ &  $\omega$ &  $\bar{\omega}$ &  $\omega$ &  $\bar{\omega}$ &  $1$ \\
10 &  $1$ &  $\bar{\omega}$ &  $\bar{\omega}$ &  $1$ &  $\omega$ &  $\omega$ \\
11 &  $1$ &  $\bar{\omega}$ &  $\omega$ &  $\omega$ &  $1$ &  $\bar{\omega}$ \\
12 &  $1$ &  $\bar{\omega}$ &  $\bar{\omega}$ &  $\bar{\omega}$ &  $1$ &  $1$ \\
13 &  $1$ &  $\bar{\omega}$ &  $1$ &  $1$ &  $\bar{\omega}$ &  $\bar{\omega}$ \\
14 &  $1$ &  $\omega$ &  $1$ &  $\omega$ &  $1$ &  $\omega$ \\
15 &  $1$ &  $\omega$ &  $\omega$ &  $1$ &  $\omega$ &  $1$ \\
16 &  $1$ &  $\omega$ &  $\omega$ &  $1$ &  $1$ &  $\omega$ \\
17 &  $1$ &  $1$ &  $\omega$ &  $\omega$ &  $\omega$ &  $1$ \\
18 &  $1$ &  $1$ &  $\bar{\omega}$ &  $1$ &  $\bar{\omega}$ &  $\bar{\omega}$ \\
19 &  $1$ &  $1$ &  $\bar{\omega}$ &  $\bar{\omega}$ &  $1$ &  $\bar{\omega}$ \\
20 &  $1$ &  $1$ &  $1$ &  $\omega$ &  $\omega$ &  $\omega$ \\
21 &  $1$ &  $\bar{\omega}$ &  $1$ &  $\bar{\omega}$ &  $\bar{\omega}$ &  $1$ \\
\end{tabular}}
\qquad\qquad\qquad
\subfloat[][]{
\label{18_9_Bases}
\begin{tabular}{cccc}
1 & 2 & 17 & 18 \\
1 & 3 & 14 & 15 \\
2 & 3 & 11 & 12 \\
4 & 7 & 13 & 15 \\
4 & 8 & 12 & 16 \\
5 & 6 & 13 & 14 \\
5 & 9 & 11 & 16 \\
6 & 9 & 10 & 18 \\
7 & 8 & 10 & 17 \\
\end{tabular}}
\qquad
\subfloat[][]{\label{21_7_Bases}
\begin{tabular}{cccccc}
1 & 2 & 3 & 4 & 5 & 6 \\
1 & 7 & 8 & 9 & 10 & 11 \\
2 & 7 & 12 & 13 & 14 & 15 \\
3 & 8 & 12 & 16 & 17 & 18 \\
4 & 9 & 13 & 16 & 19 & 20 \\
5 & 10 & 14 & 17 & 19 & 21 \\
6 & 11 & 15 & 18 & 20 & 21 \\
\end{tabular}}
     \caption{\protect\subref{18_9}: The 18 indexed rank-1 projectors (shown as kets) in the $18^1_2 - 9_4^4$ KS set \cite{Cabello18_9}, with an overbar denoting a negative sign.  \protect\subref{18_9_Bases}: The 9 complete contexts in the $18^1_2 - 9_4^4$ KS set, using the projectors from \protect\subref{18_9}.   \protect\subref{21_7}: The 21 indexed rank-1 projectors in the $21^1_2 - 7_6^6$ KS set \cite{lisonvek2014kochen}, with $\omega = e^{i 2 \pi/3}$ and $\bar{\omega} = e^{-i 2 \pi/3}$.  \protect\subref{21_7_Bases}: The 7 complete contexts in the $21^1_2 - 7_6^6$ KS set, using the projectors from \protect\subref{21_7}.}\label{D46}
\end{figure}

\textbf{The Simplest KS Sets}:---  Here we review the simplest known KS sets in $d=3,4,6,8$, which we will use as seeds to generate sets in all higher dimensions.  We introduce the compact symbol $R-B$ to denote a KS set with $R$ distinct projectors (rays) and $B$ complete contexts (bases).  In general, the projectors in a given set may be of different ranks $r$ and each may occur in a different number of contexts --- which we call the {\it multiplicity} $m$ of the projector.  We can separate the projectors in the set into classes using these properties, and in the more detailed symbol we give a sequence of $R_m^r$ values to denote the number, $R$, of projectors of rank $r$ and multiplicity $m$.  Because the projectors may be of different rank, the number $c$ of projectors in a complete context may be less than the dimension $d$, and we also separate the contexts into classes by giving a sequence of $B_c^d$ values to indicate that there are $B$ contexts of $c$ projectors each (with the Hilbert space dimension $d$ occurring as a superscript in every such symbol).  It is easy to look at these symbols and check the parity property of a set, since all the $m$ must be even, and $B$ (which is the sum of the individual $B$s) must be odd.

One of the simplest known KS sets \cite{Cabello18_9, Peres24, Mermin_SquareStar} has the compact symbol $18-9$ and detailed symbol $18^1_2 - 9_4^4$, showing that it contains 18 rank-1 projectors, each occurring twice among 9 complete contexts in $d=4$, with each context containing 4 projectors.  The detailed form of the rays and bases in this set is shown in Figs. \ref{18_9} and \ref{18_9_Bases}.  Another simple KS set \cite{lisonvek2014kochen} has the compact symbol $21-7$ and the detailed symbol $21^1_2 - 7_6^6$, and its projectors and contexts are shown in Figs. \ref{21_7} and \ref{21_7_Bases}.

 A rank-$r$ projector $\Pi^r$ is an $r$-dimensional subspace, which we represent with a spanning set of $r$ mutually orthogonal rank-1 projectors $\{|e_i\rangle\langle e_i|\}$, such that $\Pi^r = \sum_{i=1}^r |e_i\rangle\langle e_i|$.  There are $r(r-1)$ internal degrees of freedom in choosing the representative set for a rank-$r$ projector.  For example, given an orthogonal pair, $|0\rangle$ and $|1\rangle$, that spans the subspace of a rank-2 projector, any other pair $|e_1\rangle = \cos(\theta/2)|0\rangle + \sin(\theta/2)e^{i \phi} |1\rangle$ and $|e_2\rangle = \sin(\theta/2)|0\rangle - \cos(\theta/2)e^{i \phi} |1\rangle$ spans it as well, and thus there are two internal degrees of freedom ($\theta$ and $\phi$) in choosing a representative pair of rank-1 projectors.  Regardless of its rank, each projector represents a single yes/no test in an experiment, and thus there is practical value in finding the KS set with the smallest number of general-rank projectors.


The next KS set \cite{waegell2015parity}, shown in Figs. \ref{34_9} and \ref{34_9_Bases}, has the compact symbol $30-9$ and detailed symbol $4^2_2 2^1_4 24_2^1 - 8_7^8 1_8^8$, showing that it has 4 rank-2 projectors of multiplicity 2, 2 rank-1 projectors of multiplicity 4, and 24 rank-1 projectors of multiplicity 2, which form 8 contexts of 7 projectors and 1 context of 8 projectors in $d=8$.  One can obtain an all-rank-1 $34-9$ set with detailed symbol $2^1_4 32^1_2 - 9_8^8$ set by reinterpreting each of the rank-2 projectors, shown in boldface in Fig. \ref{34_9_Bases}, as any pair of rank-1 projectors that satisfy $\Pi^2 = |e_1\rangle\langle e_1| + |e_2\rangle\langle e_2|$.


\begin{figure}
\centering
     \subfloat[][]{\label{34_9}
     \scalebox{0.7}{
\begin{tabular}{c|cccccccccc|cccccccc}
1 &  $1$ &  $\bar{1}$ &  $0$ &  $0$ &  $0$ &  $0$ &  $0$ &  $0$ & & 18 &  $0$ &  $1$ &  $0$ &  $0$ &  $0$ &  $0$ &  $0$ &  $1$ \\
2 &  $0$ &  $0$ &  $1$ &  $\bar{1}$ &  $0$ &  $0$ &  $0$ &  $0$ & &19 &  $0$ &  $1$ &  $0$ &  $0$ &  $0$ &  $0$ &  $0$ &  $\bar{1}$  \\
3 &  $1$ &  $0$ &  $1$ &  $0$ &  $0$ &  $0$ &  $0$ &  $0$ & & 20 &  $0$ &  $0$ &  $1$ &  $0$ &  $0$ &  $1$ &  $0$ &  $0$\\
4 &  $1$ &  $0$ &  $\bar{1}$ &  $0$ &  $0$ &  $0$ &  $0$ &  $0$ & & 21 &  $0$ &  $0$ &  $0$ &  $1$ &  $0$ &  $0$ &  $1$ &  $0$ \\
5 &  $1$ &  $\bar{1}$ &  $1$ &  $\bar{1}$ &  $1$ &  $\bar{1}$ &  $1$ &  $1$ & & 22 &  $0$ &  $0$ &  $0$ &  $1$ &  $0$ &  $0$ &  $\bar{1}$ &  $0$  \\
6 &  $1$ &  $\bar{1}$ &  $\bar{1}$ &  $1$ &  $1$ &  $1$ &  $\bar{1}$ &  $1$ & &23 &  $0$ &  $0$ &  $0$ &  $0$ &  $1$ &  $0$ &  $1$ &  $0$  \\
7 &  $1$ &  $\bar{1}$ &  $1$ &  $\bar{1}$ &  $1$ &  $\bar{1}$ &  $\bar{1}$ &  $\bar{1}$ & & 24 &  $0$ &  $0$ &  $0$ &  $0$ &  $1$ &  $0$ &  $\bar{1}$ &  $0$\\
8 &  $1$ &  $\bar{1}$ &  $\bar{1}$ &  $1$ &  $1$ &  $1$ &  $1$ &  $\bar{1}$ & & 25 &  $0$ &  $0$ &  $0$ &  $0$ &  $0$ &  $1$ &  $0$ &  $1$ \\
9 &  $0$ &  $1$ &  $0$ &  $0$ &  $1$ &  $0$ &  $0$ &  $0$ & & 26 &  $0$ &  $0$ &  $0$ &  $0$ &  $0$ &  $1$ &  $0$ &  $\bar{1}$ \\
10 &  $0$ &  $0$ &  $0$ &  $1$ &  $0$ &  $\bar{1}$ &  $0$ &  $0$ & & 27 &  $1$ &  $1$ &  $1$ &  $1$ &  $1$ &  $1$ &  $1$ &  $\bar{1}$ \\
11 &  $1$ &  $0$ &  $0$ &  $0$ &  $0$ &  $0$ &  $0$ &  $1$ & & 28 &  $1$ &  $1$ &  $1$ &  $1$ &  $1$ &  $1$ &  $\bar{1}$ &  $1$\\
12 &  $1$ &  $0$ &  $0$ &  $0$ &  $0$ &  $0$ &  $0$ &  $\bar{1}$ & & 29 &  $1$ &  $1$ &  $1$ &  $1$ &  $\bar{1}$ &  $\bar{1}$ &  $1$ &  $\bar{1}$\\
13 &  $0$ &  $1$ &  $0$ &  $0$ &  $\bar{1}$ &  $0$ &  $0$ &  $0$ & & 30 &  $1$ &  $1$ &  $1$ &  $1$ &  $\bar{1}$ &  $\bar{1}$ &  $\bar{1}$ &  $1$\\
14 &  $0$ &  $0$ &  $1$ &  $0$ &  $0$ &  $0$ &  $1$ &  $0$ & & 31 &  $1$ &  $1$ &  $\bar{1}$ &  $\bar{1}$ &  $1$ &  $\bar{1}$ &  $1$ &  $1$\\
15 &  $0$ &  $0$ &  $1$ &  $0$ &  $0$ &  $0$ &  $\bar{1}$ &  $0$ & & 32 &  $1$ &  $1$ &  $\bar{1}$ &  $\bar{1}$ &  $1$ &  $\bar{1}$ &  $\bar{1}$ &  $\bar{1}$\\
16 &  $0$ &  $0$ &  $0$ &  $1$ &  $0$ &  $1$ &  $0$ &  $0$ & & 33 &  $1$ &  $1$ &  $\bar{1}$ &  $\bar{1}$ &  $\bar{1}$ &  $1$ &  $1$ &  $1$\\
17 &  $1$ &  $0$ &  $0$ &  $0$ &  $\bar{1}$ &  $0$ &  $0$ &  $0$ & & 34 &  $1$ &  $1$ &  $\bar{1}$ &  $\bar{1}$ &  $\bar{1}$ &  $1$ &  $\bar{1}$ &  $\bar{1}$\\
\end{tabular}}}
\qquad
\subfloat[][]{\label{34_9_Bases}
\scalebox{0.9}{
\begin{tabular}{cccccccc}
\textbf{1} & \textbf{2} & 23 & 26 & 28 & 29 & 32 & 33 \\
\textbf{1} & \textbf{2} & 24 & 25 & 27 & 30 & 31 & 34 \\
\textbf{3} & \textbf{4} & 9 & 13 & 21 & 22 & 25 & 26 \\
\textbf{3} & \textbf{4} & 10 & 16 & 18 & 19 & 23 & 24 \\
\textbf{5} & \textbf{6} & 9 & 12 & 20 & 21 & 30 & 33 \\
\textbf{5} & \textbf{6} & 10 & 15 & 17 & 18 & 27 & 32 \\
\textbf{7} & \textbf{8} & 9 & 11 & 20 & 22 & 29 & 34 \\
\textbf{7} & \textbf{8} & 10 & 14 & 17 & 19 & 28 & 31 \\
9 & 10 & 11 & 12 & 13 & 14 & 15 & 16 \\
\end{tabular}}}
     \caption{\protect\subref{34_9}: The 34 indexed rank-1 projectors (shown as kets) of the $2^1_4 32^1_2 -9^8_8$ KS set \cite{waegell2015parity}, with an overbar denoting a negative sign.  \protect\subref{34_9_Bases}: The 9 complete contexts in the $2^1_4 32^1_2 - 9_8^8$ KS set, comprised of the projectors of \protect\subref{34_9}.  If the pairs of rank-1 projectors indexed $j$ and $j+1$ are regarded as rank-2 projectors for $j=1,3,5,7$ (shown in boldface in \protect\subref{34_9_Bases}), this set can be reinterpreted as a $4^2_2 2^1_4 24^1_2-8^8_7 1^8_8$.}\label{D8}
\end{figure}

Although we do not use them in this letter, we complete our survey of the smallest known KS sets with the set of 31 projectors given by Kochen and Conway and the set of 33 projectors given by Peres \cite{peres1993quantum}, both in dimension $d=3$.  Both of these KS sets involve incomplete contexts --- meaning that the third projector of the context is not included in the original set.  Since this letter only deals with KS sets involving complete contexts, we add projectors to each set in order to complete all of the original contexts.  The 33 rays of Peres then give a critical $57-40$ set, while the 31 rays of Kochen and Conway give a noncritical $61-46$ set, which we reduced to a critical $49-36$ set using a computational search.  These sets are detailed in the Supplemental Information.

\textbf{Methods of Generating New KS Sets}:---  Next we review and improve upon several known methods of constructing a new KS set using one or more known KS sets in lower dimensions.

{\it Improved Penrose-Zimba method}:--- This method was introduced by Penrose and Zimba, who showed that two KS sets $R_1 - B_1$ in dimension $d_1$ and $R_2 - B_2$ in $d_2$ can be combined to give a new KS set $R-B$ in dimension $d=d_1 + d_2$ with $R = R_1 + R_2$ and $B = B_1 B_2 $.  If either of $d_1$ or $d_2$ is odd, then this set is critical.  However, we show that if both $R_1 - B_1$ and $R_2 - B_2$ are parity sets --- which only exist in even dimensions --- then we can construct a critical KS set with $B =$ max$\{B_1,B_2\}$.  For simplicity, we let $B_1 \leq B_2$, so that $B=B_2$.

To begin, we construct the set $\{\Pi_1 \}$ by appending $d_2$ zeros to the end of each ket in $R_1-B_1$ in order to promote it to dimension $d$.  Similarly, we construct the set $\{\Pi_2\}$ by appending $d_1$ zeros to the beginning of each ket from $R_2-B_2$.  Together these are the $R$ projectors of the new KS set of the standard Penrose-Zimba method.

All of the projectors in $\{\Pi_1\}$ are orthogonal to all the projectors in $\{\Pi_2\}$, while each subset inherits the internal pattern of orthogonality relations from its parent KS set.  Importantly, this means that the set of $d_1$ projectors from within $\{\Pi_1\}$ corresponding to any one complete context in $R_1-B_1$, along with the $d_2$ projectors from within $\{\Pi_2\}$ corresponding to any one complete context in $R_2-B_2$, automatically comprise a complete context in dimension $d$.

We explicitly construct the contexts in $R-B$ by pairing off each context in $R_1-B_1$ with a different context from $R_2-B_2$, finally pairing the remaining contexts in $R_2-B_2$ with $\delta_B = B_2-B_1$ extra copies of any one of the contexts in $R_1-B_1$.  The order in which they are paired is arbitrary, although it can affect the ranks of the projector in the resulting set.

Because both $B_1$ and $B_2$ are odd in a parity set, $\delta_B$ is even, and thus the multiplicities of the projectors in the copied context always increase in even increments, ensuring the parity of the resulting KS set $R-B$.  Note that we might also use copies of several different contexts from $R_1 - B_1$, provided that each is always used an even number of times; such variations can give rise to KS sets whose projectors have different multiplicities.  The criticality of the larger set $R_2-B_2$ also guarantees the criticality of the derived set $R-B$.

\begin{figure}
\centering
\subfloat[][]{
\label{39_9_Bases}
\scalebox{1}{
\begin{tabular}{cccccc|cccc}
1 & 2 & 3 & 4 & 5 & 6 & {\it 1} & {\it 2} & {\it 17} & {\it 18} \\
1 & 2 & 3 & 4 & 5 & 6 & {\it 6} & {\it 9} & {\it 10} & {\it 18} \\
1 & 2 & 3 & 4 & 5 & 6 & {\it 7} & {\it 8} & {\it 10} & {\it 17} \\
1 & 9 & 11 & \textbf{7} & \textbf{8} & \textbf{10} & {\it \textbf{3}} & {\it \textbf{14}} & {\it \textbf{15}} & {\it 1} \\
2 & 12 & 14 & \textbf{7} & \textbf{13} & \textbf{15} & {\it \textbf{3}} & {\it \textbf{11}} & {\it \textbf{12}} & {\it 2} \\
3 & 12 & 18 & \textbf{8} & \textbf{16} & \textbf{17} & {\it \textbf{4}} & {\it \textbf{13}} & {\it \textbf{15}} & {\it 7} \\
4 & 9 & 19 & \textbf{13} & \textbf{16} & \textbf{20} & {\it \textbf{4}} & {\it \textbf{12}} & {\it \textbf{16}} & {\it 8} \\
5 & 14 & 19 & \textbf{10} & \textbf{17} & \textbf{21} & {\it \textbf{5}} & {\it \textbf{13}} & {\it \textbf{14}} & {\it 6} \\
6 & 11 & 18 & \textbf{15} & \textbf{20} & \textbf{21} & {\it \textbf{5}} & {\it \textbf{11}} & {\it \textbf{16}} & {\it 9} \\
\end{tabular}}}
\qquad
\subfloat[][]{
\label{Rank2}
\scalebox{1}{
\begin{tabular}{c|c}
\textbf{7} & {\it \textbf{3}} \\
\textbf{8} & {\it \textbf{15}} \\
\textbf{10} & {\it \textbf{14}} \\
\textbf{13} & {\it \textbf{12}} \\
\textbf{15} & {\it \textbf{11}} \\
\textbf{16} & {\it \textbf{4}} \\
\textbf{17} & {\it \textbf{13}} \\
\textbf{20} & {\it \textbf{16}} \\
\textbf{21} & {\it \textbf{5}} \\
\end{tabular}}}
\caption{\protect\subref{39_9_Bases}: The 9 complete contexts in the $9_2^2 6^1_4 15^1_2 - 6_{7}^{10} 3_{10}^{10}$ KS set (with the compact symbol $30-9$), starting from the projectors from Fig. \ref{18_9} (italic script) and Fig. \ref{21_7} (plain script), and with the 18 projectors in boldface comprising the 9 rank-2 projectors in this set, whose constituent projectors are shown in \protect\subref{39_9_Bases}.  If each of the rank-2 projectors is replaced by a pair of rank-1 projectors satisfying $\Pi^2 = |e_1\rangle\langle e_1| + |e_2\rangle\langle e_2|$, then this $30-9$ KS set gives rise to a $6^1_4 33^1_2 - 9_{10}^{10}$ KS set with the compact symbol $39-9$.}
\end{figure}

As an example, we apply this method to the $18_2^1-9_4^4$ \cite{Cabello18_9} and $21_2^1 - 7_6^6$ \cite{lisonvek2014kochen} sets to construct a new $6^1_4 33^1_2 - 9_{10}^{10}$ set.  The 39 rays in $d=10$ are obtained by appending 6 zeros to the beginning of those in Fig. \ref{18_9} and 4 zeros to the end of those in Fig. \ref{21_7}, as described above.  The 9 contexts of the new KS set are shown in Fig. \ref{39_9_Bases}, using italic script for the indexes of the 18 rays originating from the $18-9$ and plain script for the indexes of the 21 rays originating from the $21-7$.  Note the two extra copies of the first context from the $21-7$ that are used to pair all 9 contexts of the $18-9$.

When combining two different KS sets, some number of the projectors in the resulting KS set may comprise new higher-rank projectors --- depending on the order in which the contexts are paired.  Specifically, if all $m$ contexts containing a particular projector from $R_1 - B_1$ are paired with all $m$ contexts containing a particular projector from $R_2 - B_2$, then those two projectors combine into a single projector in the new set whose rank is the sum of their two ranks.  In general, this means that the new $R-B$ KS set may be reduced to a more compact $R'-B$ set, with $R'< R$.

The bases from the two parent sets in our example have been paired off in an optimal way that gives rise to 9 rank-2 projectors within the set, and thus the $39-9$ can be reduced to a $9_2^2 6^1_4 15^1_2 - 6_{7}^{10} 3_{10}^{10}$ with the compact symbol $30-9$.

{\it Rank-Scaling method}:--- In what follows, we consider the simplified case in which all projectors in the parent set are rank-1, but this reasoning can be easily generalized to accommodate general-rank projectors.  When combining two copies of the same $R-B$ KS set of dimension $d$ using the improved Penrose-Zimba method, the two copies are put in orthogonal subspaces of a $2d$-dimensional Hilbert space.  Therefore each of the $R$ projectors in the original set generates two orthogonal rank-1 projectors, and together these two form a single rank-2 projector in the new set.  The $B$ original contexts in dimension $d$ then give rise to $B$ contexts in dimension $2d$, each comprised of the $d$ rank-2 projectors in the same way that the original set was comprised of the $d$ parent rank-1 projectors.  If a third copy of the original KS set is added in yet another mutually orthogonal subspace, we end up with rank-3 projectors in a $3d$-dimensional Hilbert space, and so on.  Therefore an $R-B$ KS set in dimension $d$ gives rise to $R-B$ KS sets in all dimensions $nd$, with the ranks of the $R$ projectors scaled by a factor $n$, and the structure of the $B$ contexts unchanged (i.e., with all projectors replaced by their rank-scaled counterparts).




{\it Cabello, Estebaranz, and Garc{\'\i}a-Alcaine method}:--- Another method, given by Cabello, et al. \cite{cabello2005recursive}, can be used to obtain new KS sets $R'-B'$ in dimension $d'$ from a known set $R-B$ in dimension $d$, with $d < d' < 2d$, $R'\leq 2R + 3'$ and $B'\leq 2B+1$, although in general the new KS set is neither critical nor a parity set.  In order to obtain the new set we first take the original $R-B$ set, and append $\delta= d'-d$ zeros to the end of each ket to construct $R$ $d'$-dimensional projectors $\{\Pi_1\}$.  Likewise, we construct $R$ projectors $\{\Pi_2\}$ by appending $\delta$ zeros to the beginning of the kets from $R-B$.  Next, we define the rank-$\delta$ projector $\Pi_l$ onto the first $\delta$ dimensions of the new $d'$-dimensional space, the rank-$\delta$ projector $\Pi_r$ onto the last $\delta$ dimensions, and the rank-$\rho$ projector $\Pi_c$ onto the center $\rho=2d-d'$ dimensions, such that these three projectors comprise a complete context.  The projectors of $\{\Pi_1\}$ together with $\Pi_r$ comprise $B$ more contexts, and $\{\Pi_2\}$ together with $\Pi_l$ comprise yet $B$ more.  This gives rise to a KS set with $B'= 2B+1$ contexts, which is not critical in general.

{\it Improved Matsuno method}:--- Matsuno \cite{matsuno2007construction} refined the Cabello et al. method to obtain the explicit lower bound of $R'\leq 2R-1$ or better, which depend on the details of the all-rank-1 parent set $R-B$.  In order to apply this method to KS sets with projectors of arbitrary rank, we choose a particular decomposition of each higher-rank projector in terms of rank-1 projectors, which usually results in some groups of rank-1 projectors that comprise higher-rank projectors in the new KS set.  The Matsuno method only works if some specific subset $V$ of $\delta$ projectors from $R-B$ comprise an orthonormal basis for the first $\delta$ dimensions of the $d'$-dimensional space, and thus some unitary may have to be applied to the entire set $R-B$ to put one context into a suitable form.  To apply this method, we begin by constructing the set of $R$ projectors $\{\Pi_1\}$, just as in the Cabello et al. method.  Next we define the block transformation matrix in $d'$ dimensions,
\begin{equation}
T = \left(
  \begin{array}{ccc}
    0 & 0 & I^\delta \\
    0 & I^\rho & 0 \\
    I^\delta & 0 & 0 \\
  \end{array}
\right),
\end{equation}
which swaps the first $\delta$ dimensions of the new space with the last $\delta$ dimensions, where $I^\delta$ is the identity matrix in dimension $\delta$ and $I^\rho$ is the identity in dimension $\rho$.  Then we construct the set of $R$ projectors $\{\Pi_2\} = \{T \Pi_1 T\}$, which automatically contains at least one duplicate of a projector in $\{\Pi_1\}$.  Once these duplicates have been removed, the union of $\{\Pi_1\}$ and $\{\Pi_2\}$ gives the $R'\leq 2R-1$ projectors of the new set.  In general, the complete set of contexts formed by the new set of projectors is noncritical, and must be searched to find critical subsets.

We improve the Matsuno method by giving an explicit construction for the $B' \leq 2B-1$ contexts of a critical $R'-B'$ KS set, without the need for a search.  We label the subsets of new projectors corresponding to the orthonormal subset $V$ in $R-B$ as $\{\Pi_1^V\}$ and $\{\Pi_2^V\}$ --- respectively before and after the action of $T$.  Each of the $B$ $d$-dimensional contexts in $R-B$ that appear within $\{\Pi_1\}$ can be promoted to a $d'$-dimensional context by adding the subset $\{\Pi_2^V\}$, and likewise each of the $B$ contexts that appear within $\{\Pi_2\}$ can be promoted by adding the subset $\{\Pi_1^V\}$, producing $2B$ contexts in total.  Now, when the duplicate projectors are removed from $\{\Pi_2\}$, their labels within these $2B$ contexts are replaced with the corresponding labels from $\{\Pi_1\}$, which causes at least one context in the set to become a duplicate of another.  Removing these duplicates leaves the $B' \leq 2B-1$ contexts that comprise a critical $R'-B'$ KS set.

\textbf{Results}:---  Applying the methods of the previous sections to the simplest known KS sets in dimensions $d=4,6,8$ allows us to construct the simplest known KS sets in all dimensions $d\geq5$ \footnote{The $29-16$ KS set in $d=5$ was originally discovered by Cabello et al. \cite{cabello2005recursive} --- see the Supplemental Information} --- which are parity sets for all even $d$.  In Fig. \ref{Hierarchy} we show the numbers of projectors in the smallest KS sets obtained by these methods for all dimensions, using either general-rank projectors ($R^r-B$), or all-rank-1 projectors ($R^1-B$) as alternative standards for `smallest,' since these are optimal for different applications.

\begin{figure}
\centering
\caption{Simplest known KS sets with the compact symbol $R-B$ in dimension $d$, with integers $n,l\geq1$ and $m\geq2$.  The simplest known KS sets using projectors of any combination of ranks are listed under $R^r-B$ , while the simplest using only rank-1 projectors are listed under $R^1-B$.  These sets are obtained by some combination of the improved Penrose-Zimba method, the rank-scaling method, the Cabello et al. method and the Matsuno method, as explained in the text.  All of these sets are critical, except for the last two rows, for which the smallest critical subsets are unknown. There are redundancies and alternative choices throughout this list.  Explicit examples of these KS sets for all $3 \leq d\leq 11$ are given in the text for even $d$ and in the Supplemental Information for odd $d$.}\label{Hierarchy}
\scalebox{1}{
\begin{tabular}{c|c|c}
  $d$ & $R^r-B$ & $R^1-B$ \\
  \hline
  $3n$ & $49-36$ & $49n-36$ \\
  $4n$ & $18-9$ & $18n-9$ \\
  $5n$ & $29-16$ & $29n-16$ \\
  $6n$ & $21-7$ & $21n-7$ \\
  $7n$ & $32-12$ & $32n-12$ \\
  $8n$ & $18-9$ & $34n-9$ \\
  $9n$ & $39-13$ & $39n-13$ \\
  $10n$ & $30-9$ & $39n-9$ \\
  $11n$ & $40-12$ & $40n-12$ \\
  $6m+1$ &$43-12$ & $(21m+11)-12$ \\
  $6n+2$ & & $(21n+13)-9$ \\
   $6m+3$ & $57-13$ & $(21m+18)-13$ \\
  $6n+4$ & & $(21n+18)-9$ \\
   $6m+5$ & $61-13$ & $(21m+20)-13$ \\
  $6n + 4l$ & $30-9$ &  \\
  $2n+5$ & $45-15$ &  \\
  $2n+3$ & $39-19$ &  \\
\end{tabular}}
\end{figure}

Note that $18-9$, $21-7$, and $30-9$ are the only entries that appear in the $R^r-B$ column of Fig. \ref{Hierarchy} for even dimensions, because for integers $l,n\geq1$, any dimension $4l$ has $18-9$ KS sets, any dimension $6n$ has $21-7$ KS sets, and any other even dimension can be obtained as $d=4l+6n$, giving $30-9$ KS sets using the improved Penrose-Zimba method.

For odd dimensions, we apply the improved Matsuno method to the $21n-7$ KS sets in dimension $d=6n$ to obtain the entries $43-12$, $57-13$, $61-13$ in the $R^r-B$ column of Fig. \ref{Hierarchy} for odd dimensions.  Notice that in $d=11$, the original Matsuno method followed by a computational search yields a smaller critical $40-12$ KS set than the improved method ($41-13$); however complete searches quickly become computationally intractable as the dimension increases, and thus the smallest KS sets that can be obtained by this method are not known for $d \geq 13$.

If we prefer to minimize the number of projectors over the number of contexts, the original Cabello, et al. method applied to the $21n-7$ KS sets gives generally noncritical $45-15$ KS sets in all odd dimensions $d\geq7$, and applied to the $18n-9$ KS sets in $d=4n$ gives noncritical $39-19$ KS sets in all odd dimensions $d\geq 5$.

These methods also produce all-rank-1 KS sets in all dimensions $d\geq9$, which are fundamentally built around the $21n-7$ KS sets in $d=6n$, either combined with other seed sets in dimensions $d=4,8$ using the improved Penrose-Zimba method, or generalized to odd dimensions $d=6n+1,3,5$ using the improved Matsuno method.  As a result, the maximum number of rank-1 projectors in any such KS set scales as $R \approx 7d/2$ for large $d$, which improves on the previously known scaling of $R\approx 9d/2$ \cite{zimba1993bell, cabello2005recursive, yu2011minimal}.


Finally, our results reveal several aspects of the hierarchy of KS sets that were not previously known.  Most importantly, the number of complete contexts in a minimal KS set does not scale with dimension $d$, and we have shown that these sets require at most $7$ or $9$ contexts in any even dimension and $12$ or $13$ contexts in any odd dimension.  Next, the number of general-rank projectors in a minimal KS set also does not scale with $d$, there being at most $18$, $21$, or $30$ projectors in any even dimension and $39$ in any odd dimension.

We hope that the new KS sets and methods we have presented, combined with the existing ones we have reviewed, give an up-to-date picture of the state of the art.  Beyond their foundational significance for tests of the Bell-Kochen-Specker theorem or quantum nonlocality \cite{Bell1,EPR}, the general hierarchy of minimal KS sets presented here should also be ideally suited for a number of proposed applications of KS sets, including quantum computation \cite{galvao2005discrete, spekkens2008negativity, howard2014contextuality, bermejo2016contextuality}, quantum key distribution \cite{ekert1991quantum, shor2000simple, singh2016quantum}, parity oblivious transfer \cite{bennett1992practical}, random number generation \cite{abbott2012strong}, quantum dimension certification \cite{CabelloKSDim}, and relational database theory \cite{abramsky2013relational}.

\textbf{Acknowledgments}:---  We are pleased to present this letter in celebration of the 50th anniversary of the Bell-Kochen-Specker theorem.  This research was supported (in part) by the Fetzer-Franklin Fund  of  the  John  E. Fetzer  Memorial  Trust.

\bibliographystyle{ieeetr}
\bibliography{KS_All_D.bbl}

\end{document}


\title{Supplemental Information for\\ ``The Minimum Complexity of Kochen-Specker Sets Does Not Scale with Dimension''}

\author{Mordecai Waegell and P.K. Aravind}
\maketitle

\begin{figure}
\centering
\subfloat[][]{
\label{D5_29_16}
\scalebox{0.9}{
\begin{tabular}{c|ccccc}
1 &  $1$ &  $0$ &  $0$ &  $0$ &  $0$ \\
2 &  $0$ &  $1$ &  $0$ &  $0$ &  $0$ \\
3 &  $0$ &  $0$ &  $1$ &  $0$ &  $0$ \\
4 &  $1$ &  $1$ &  $1$ &  $1$ &  $0$ \\
5 &  $1$ &  $1$ &  $1$ &  $\bar{1}$ &  $0$ \\
6 &  $1$ &  $\bar{1}$ &  $1$ &  $1$ &  $0$ \\
7 &  $1$ &  $\bar{1}$ &  $1$ &  $\bar{1}$ &  $0$ \\
8 &  $1$ &  $\bar{1}$ &  $\bar{1}$ &  $1$ &  $0$ \\
9 &  $1$ &  $\bar{1}$ &  $\bar{1}$ &  $\bar{1}$ &  $0$ \\
10 &  $1$ &  $1$ &  $0$ &  $0$ &  $0$ \\
11 &  $1$ &  $0$ &  $0$ &  $1$ &  $0$ \\
12 &  $1$ &  $0$ &  $0$ &  $\bar{1}$ &  $0$ \\
13 &  $1$ &  $0$ &  $\bar{1}$ &  $0$ &  $0$ \\
14 &  $0$ &  $1$ &  $0$ &  $1$ &  $0$ \\
15 &  $0$ &  $1$ &  $0$ &  $\bar{1}$ &  $0$ \\
16 &  $0$ &  $1$ &  $\bar{1}$ &  $0$ &  $0$ \\
17 &  $0$ &  $0$ &  $1$ &  $1$ &  $0$ \\
18 &  $0$ &  $0$ &  $1$ &  $\bar{1}$ &  $0$ \\
19 &  $0$ &  $0$ &  $0$ &  $0$ &  $1$ \\
20 &  $0$ &  $1$ &  $1$ &  $1$ &  $1$ \\
21 &  $0$ &  $1$ &  $1$ &  $\bar{1}$ &  $1$ \\
22 &  $0$ &  $1$ &  $\bar{1}$ &  $\bar{1}$ &  $\bar{1}$ \\
23 &  $0$ &  $1$ &  $\bar{1}$ &  $1$ &  $\bar{1}$ \\
24 &  $0$ &  $1$ &  $1$ &  $\bar{1}$ &  $\bar{1}$ \\
25 &  $0$ &  $1$ &  $1$ &  $1$ &  $\bar{1}$ \\
26 &  $0$ &  $1$ &  $0$ &  $0$ &  $1$ \\
27 &  $0$ &  $0$ &  $0$ &  $1$ &  $1$ \\
28 &  $0$ &  $0$ &  $0$ &  $1$ &  $\bar{1}$ \\
29 &  $0$ &  $0$ &  $1$ &  $0$ &  $\bar{1}$ \\
\end{tabular}}}
\qquad
\subfloat[][]{
\label{D5_29_16_Bases}
\scalebox{1}{
\begin{tabular}{ccccc}
1 & 2 & 3 & 27 & 28 \\
1 & 2 & 17 & 18 & 19 \\
1 & 3 & 14 & 15 & 19 \\
1 & 14 & 21 & 22 & 29 \\
1 & 15 & 20 & 23 & 29 \\
1 & 16 & 20 & 24 & 28 \\
1 & 16 & 21 & 25 & 27 \\
1 & 17 & 23 & 24 & 26 \\
1 & 18 & 22 & 25 & 26 \\
2 & 3 & 11 & 12 & 19 \\
4 & 7 & 13 & 15 & 19 \\
4 & 8 & 12 & 16 & 19 \\
5 & 6 & 13 & 14 & 19 \\
5 & 9 & 11 & 16 & 19 \\
6 & 9 & 10 & 18 & 19 \\
7 & 8 & 10 & 17 & 19 \\
\end{tabular}}}
\qquad
\subfloat[][]{
\label{D7_32_12}
\scalebox{0.8}{
\begin{tabular}{c|ccccccc}
1 &  $1$ &  $0$ &  $0$ &  $0$ &  $0$ &  $0$ &  $0$ \\
2 &  $0$ &  $1$ &  $0$ &  $0$ &  $0$ &  $0$ &  $0$ \\
3 &  $0$ &  $0$ &  $1$ &  $0$ &  $0$ &  $0$ &  $0$ \\
4 &  $0$ &  $0$ &  $0$ &  $1$ &  $0$ &  $0$ &  $0$ \\
5 &  $0$ &  $0$ &  $0$ &  $0$ &  $1$ &  $0$ &  $0$ \\
6 &  $0$ &  $0$ &  $0$ &  $0$ &  $0$ &  $1$ &  $0$ \\
7 &  $0$ &  $0$ &  $1$ &  $1$ &  $\omega^2$ &  $\omega^2$ &  $0$ \\
8 &  $0$ &  $1$ &  $0$ &  $\omega^2$ &  $\omega^2$ &  $1$ &  $0$ \\
9 &  $0$ &  $1$ &  $\omega^2$ &  $0$ &  $1$ &  $\omega^2$ &  $0$ \\
10 &  $0$ &  $1$ &  $1$ &  $\omega^4$ &  $0$ &  $\omega^4$ &  $0$ \\
11 &  $0$ &  $1$ &  $\omega^4$ &  $1$ &  $\omega^4$ &  $0$ &  $0$ \\
12 &  $1$ &  $0$ &  $0$ &  $\omega^5$ &  $-1$ &  $\omega^5$ &  $0$ \\
13 &  $1$ &  $0$ &  $\omega^5$ &  $0$ &  $\omega^5$ &  $-1$ &  $0$ \\
14 &  $1$ &  $0$ &  $\omega$ &  $-1$ &  $0$ &  $\omega$ &  $0$ \\
15 &  $1$ &  $0$ &  $-1$ &  $\omega$ &  $\omega$ &  $0$ &  $0$ \\
16 &  $1$ &  $-1$ &  $0$ &  $0$ &  $\omega$ &  $\omega$ &  $0$ \\
17 &  $1$ &  $\omega$ &  $0$ &  $\omega$ &  $0$ &  $-1$ &  $0$ \\
18 &  $1$ &  $\omega^5$ &  $0$ &  $-1$ &  $\omega^5$ &  $0$ &  $0$ \\
19 &  $1$ &  $\omega^5$ &  $-1$ &  $0$ &  $0$ &  $\omega^5$ &  $0$ \\
20 &  $1$ &  $\omega$ &  $\omega$ &  $0$ &  $-1$ &  $0$ &  $0$ \\
21 &  $1$ &  $-1$ &  $\omega^5$ &  $\omega^5$ &  $0$ &  $0$ &  $0$ \\
22 &  $0$ &  $0$ &  $0$ &  $0$ &  $0$ &  $0$ &  $1$ \\
23 &  $0$ &  $0$ &  $0$ &  $1$ &  $\omega^4$ &  $1$ &  $\omega$ \\
24 &  $0$ &  $0$ &  $1$ &  $0$ &  $1$ &  $\omega^4$ &  $\omega$ \\
25 &  $0$ &  $0$ &  $1$ &  $\omega^2$ &  $0$ &  $1$ &  $\omega^5$ \\
26 &  $0$ &  $0$ &  $1$ &  $\omega^4$ &  $\omega^4$ &  $0$ &  $-1$ \\
27 &  $0$ &  $1$ &  $0$ &  $0$ &  $\omega^4$ &  $\omega^4$ &  $-1$ \\
28 &  $0$ &  $1$ &  $0$ &  $1$ &  $0$ &  $\omega^2$ &  $\omega^5$ \\
29 &  $0$ &  $1$ &  $0$ &  $\omega^4$ &  $1$ &  $0$ &  $\omega$ \\
30 &  $0$ &  $1$ &  $\omega^4$ &  $0$ &  $0$ &  $1$ &  $\omega$ \\
31 &  $0$ &  $1$ &  $1$ &  $0$ &  $\omega^2$ &  $0$ &  $\omega^5$ \\
32 &  $0$ &  $1$ &  $\omega^2$ &  $\omega^2$ &  $0$ &  $0$ &  $-1$ \\
\end{tabular}}}
\qquad
\subfloat[][]{
\label{D7_32_12_Bases}
\scalebox{1}{
\begin{tabular}{ccccccc}
1 & 2 & 3 & 4 & 5 & 6 & 22 \\
1 & 2 & 7 & 23 & 24 & 25 & 26 \\
1 & 3 & 8 & 23 & 27 & 28 & 29 \\
1 & 4 & 9 & 24 & 27 & 30 & 31 \\
1 & 5 & 10 & 25 & 28 & 30 & 32 \\
1 & 6 & 11 & 26 & 29 & 31 & 32 \\
1 & 7 & 8 & 9 & 10 & 11 & 22 \\
2 & 7 & 12 & 13 & 14 & 15 & 22 \\
3 & 8 & 12 & 16 & 17 & 18 & 22 \\
4 & 9 & 13 & 16 & 19 & 20 & 22 \\
5 & 10 & 14 & 17 & 19 & 21 & 22 \\
6 & 11 & 15 & 18 & 20 & 21 & 22 \\
\end{tabular}}}

\caption{\protect\subref{D5_29_16} \& \protect\subref{D5_29_16_Bases}: The 29 indexed projectors (shown as kets, where an overbar indicates a negative sign) and 16 contexts of the critical $29-16$ KS set in dimension $d=5$, obtained by applying the Matsuno method to the 4-dimensional $18-9$ KS set of Fig. 1a of the main text. \protect\subref{D7_32_12} \& \protect\subref{D7_32_12_Bases}: The 32 indexed projectors (shown as kets, with $\omega=e^{i \pi /3}$) and 12 contexts of the critical $32-12$ KS set in dimension $d=7$ obtained by applying the Matsuno method to the 6-dimensional $21-7$ KS set of Fig. 1b of the main text.}
\end{figure}

\begin{figure}
\centering
\subfloat[][]{
\label{D9_39_13}
\scalebox{0.7}{
\begin{tabular}{c|ccccccccc}
1 &  $1$ &  $0$ &  $0$ &  $0$ &  $0$ &  $0$ &  $0$ &  $0$ &  $0$ \\
2 &  $0$ &  $1$ &  $0$ &  $0$ &  $0$ &  $0$ &  $0$ &  $0$ &  $0$ \\
3 &  $0$ &  $0$ &  $1$ &  $0$ &  $0$ &  $0$ &  $0$ &  $0$ &  $0$ \\
4 &  $0$ &  $0$ &  $0$ &  $1$ &  $0$ &  $0$ &  $0$ &  $0$ &  $0$ \\
5 &  $0$ &  $0$ &  $0$ &  $0$ &  $1$ &  $0$ &  $0$ &  $0$ &  $0$ \\
6 &  $0$ &  $0$ &  $0$ &  $0$ &  $0$ &  $1$ &  $0$ &  $0$ &  $0$ \\
7 &  $0$ &  $0$ &  $1$ &  $1$ &  $\omega^2$ &  $\omega^2$ &  $0$ &  $0$ &  $0$ \\
8 &  $0$ &  $1$ &  $0$ &  $\omega^2$ &  $\omega^2$ &  $1$ &  $0$ &  $0$ &  $0$ \\
9 &  $0$ &  $1$ &  $\omega^2$ &  $0$ &  $1$ &  $\omega^2$ &  $0$ &  $0$ &  $0$ \\
10 &  $0$ &  $1$ &  $1$ &  $\omega^4$ &  $0$ &  $\omega^4$ &  $0$ &  $0$ &  $0$ \\
11 &  $0$ &  $1$ &  $\omega^4$ &  $1$ &  $\omega^4$ &  $0$ &  $0$ &  $0$ &  $0$ \\
12 &  $1$ &  $0$ &  $0$ &  $\omega^5$ &  $-1$ &  $\omega^5$ &  $0$ &  $0$ &  $0$ \\
13 &  $1$ &  $0$ &  $\omega^5$ &  $0$ &  $\omega^5$ &  $-1$ &  $0$ &  $0$ &  $0$ \\
14 &  $1$ &  $0$ &  $\omega$ &  $-1$ &  $0$ &  $\omega$ &  $0$ &  $0$ &  $0$ \\
15 &  $1$ &  $0$ &  $-1$ &  $\omega$ &  $\omega$ &  $0$ &  $0$ &  $0$ &  $0$ \\
16 &  $1$ &  $-1$ &  $0$ &  $0$ &  $\omega$ &  $\omega$ &  $0$ &  $0$ &  $0$ \\
17 &  $1$ &  $\omega$ &  $0$ &  $\omega$ &  $0$ &  $-1$ &  $0$ &  $0$ &  $0$ \\
18 &  $1$ &  $\omega^5$ &  $0$ &  $-1$ &  $\omega^5$ &  $0$ &  $0$ &  $0$ &  $0$ \\
19 &  $1$ &  $\omega^5$ &  $-1$ &  $0$ &  $0$ &  $\omega^5$ &  $0$ &  $0$ &  $0$ \\
20 &  $1$ &  $\omega$ &  $\omega$ &  $0$ &  $-1$ &  $0$ &  $0$ &  $0$ &  $0$ \\
21 &  $1$ &  $-1$ &  $\omega^5$ &  $\omega^5$ &  $0$ &  $0$ &  $0$ &  $0$ &  $0$ \\
22 &  $0$ &  $0$ &  $0$ &  $0$ &  $0$ &  $0$ &  $1$ &  $0$ &  $0$ \\
23 &  $0$ &  $0$ &  $0$ &  $0$ &  $0$ &  $0$ &  $0$ &  $1$ &  $0$ \\
24 &  $0$ &  $0$ &  $0$ &  $0$ &  $0$ &  $0$ &  $0$ &  $0$ &  $1$ \\
25 &  $0$ &  $0$ &  $0$ &  $1$ &  $\omega^2$ &  $\omega^2$ &  $0$ &  $0$ &  $1$ \\
26 &  $0$ &  $0$ &  $0$ &  $1$ &  $1$ &  $\omega^4$ &  $0$ &  $\omega^4$ &  $0$ \\
27 &  $0$ &  $0$ &  $0$ &  $0$ &  $1$ &  $\omega^2$ &  $0$ &  $1$ &  $\omega^2$ \\
28 &  $0$ &  $0$ &  $0$ &  $1$ &  $0$ &  $1$ &  $0$ &  $\omega^2$ &  $\omega^2$ \\
29 &  $0$ &  $0$ &  $0$ &  $1$ &  $\omega^4$ &  $0$ &  $0$ &  $1$ &  $\omega^4$ \\
30 &  $0$ &  $0$ &  $0$ &  $1$ &  $\omega^4$ &  $1$ &  $\omega$ &  $0$ &  $0$ \\
31 &  $0$ &  $0$ &  $0$ &  $0$ &  $1$ &  $\omega^4$ &  $\omega$ &  $0$ &  $1$ \\
32 &  $0$ &  $0$ &  $0$ &  $1$ &  $0$ &  $\omega^4$ &  $-1$ &  $0$ &  $\omega^4$ \\
33 &  $0$ &  $0$ &  $0$ &  $1$ &  $1$ &  $0$ &  $\omega^5$ &  $0$ &  $\omega^2$ \\
34 &  $0$ &  $0$ &  $0$ &  $0$ &  $1$ &  $1$ &  $\omega^5$ &  $\omega^2$ &  $0$ \\
35 &  $0$ &  $0$ &  $0$ &  $1$ &  $0$ &  $\omega^2$ &  $\omega^5$ &  $1$ &  $0$ \\
36 &  $0$ &  $0$ &  $0$ &  $1$ &  $\omega^2$ &  $0$ &  $-1$ &  $\omega^2$ &  $0$ \\
37 &  $0$ &  $0$ &  $0$ &  $0$ &  $0$ &  $1$ &  $\omega$ &  $1$ &  $\omega^4$ \\
38 &  $0$ &  $0$ &  $0$ &  $0$ &  $1$ &  $0$ &  $-1$ &  $\omega^4$ &  $\omega^4$ \\
39 &  $0$ &  $0$ &  $0$ &  $1$ &  $0$ &  $0$ &  $\omega$ &  $\omega^4$ &  $1$ \\
\end{tabular}}}
\qquad\qquad\qquad\qquad\qquad
\subfloat[][]{
\label{D11_40_12}
\scalebox{0.7}{
\begin{tabular}{c|ccccccccccc}
1 &  $1$ &  $0$ &  $0$ &  $0$ &  $0$ &  $0$ &  $0$ &  $0$ &  $0$ &  $0$ &  $0$ \\
2 &  $0$ &  $1$ &  $0$ &  $0$ &  $0$ &  $0$ &  $0$ &  $0$ &  $0$ &  $0$ &  $0$ \\
3 &  $0$ &  $0$ &  $1$ &  $0$ &  $0$ &  $0$ &  $0$ &  $0$ &  $0$ &  $0$ &  $0$ \\
4 &  $0$ &  $0$ &  $0$ &  $1$ &  $0$ &  $0$ &  $0$ &  $0$ &  $0$ &  $0$ &  $0$ \\
5 &  $0$ &  $0$ &  $0$ &  $0$ &  $1$ &  $0$ &  $0$ &  $0$ &  $0$ &  $0$ &  $0$ \\
6 &  $0$ &  $0$ &  $1$ &  $1$ &  $\omega^2$ &  $\omega^2$ &  $0$ &  $0$ &  $0$ &  $0$ &  $0$ \\
7 &  $0$ &  $1$ &  $0$ &  $\omega^2$ &  $\omega^2$ &  $1$ &  $0$ &  $0$ &  $0$ &  $0$ &  $0$ \\
8 &  $0$ &  $1$ &  $\omega^2$ &  $0$ &  $1$ &  $\omega^2$ &  $0$ &  $0$ &  $0$ &  $0$ &  $0$ \\
9 &  $0$ &  $1$ &  $1$ &  $\omega^4$ &  $0$ &  $\omega^4$ &  $0$ &  $0$ &  $0$ &  $0$ &  $0$ \\
10 &  $0$ &  $1$ &  $\omega^4$ &  $1$ &  $\omega^4$ &  $0$ &  $0$ &  $0$ &  $0$ &  $0$ &  $0$ \\
11 &  $1$ &  $0$ &  $0$ &  $\omega^5$ &  $-1$ &  $\omega^5$ &  $0$ &  $0$ &  $0$ &  $0$ &  $0$ \\
12 &  $1$ &  $0$ &  $\omega^5$ &  $0$ &  $\omega^5$ &  $-1$ &  $0$ &  $0$ &  $0$ &  $0$ &  $0$ \\
13 &  $1$ &  $0$ &  $\omega$ &  $-1$ &  $0$ &  $\omega$ &  $0$ &  $0$ &  $0$ &  $0$ &  $0$ \\
14 &  $1$ &  $0$ &  $-1$ &  $\omega$ &  $\omega$ &  $0$ &  $0$ &  $0$ &  $0$ &  $0$ &  $0$ \\
15 &  $1$ &  $-1$ &  $0$ &  $0$ &  $\omega$ &  $\omega$ &  $0$ &  $0$ &  $0$ &  $0$ &  $0$ \\
16 &  $1$ &  $\omega$ &  $0$ &  $\omega$ &  $0$ &  $-1$ &  $0$ &  $0$ &  $0$ &  $0$ &  $0$ \\
17 &  $1$ &  $\omega^5$ &  $0$ &  $-1$ &  $\omega^5$ &  $0$ &  $0$ &  $0$ &  $0$ &  $0$ &  $0$ \\
18 &  $1$ &  $\omega^5$ &  $-1$ &  $0$ &  $0$ &  $\omega^5$ &  $0$ &  $0$ &  $0$ &  $0$ &  $0$ \\
19 &  $1$ &  $\omega$ &  $\omega$ &  $0$ &  $-1$ &  $0$ &  $0$ &  $0$ &  $0$ &  $0$ &  $0$ \\
20 &  $1$ &  $-1$ &  $\omega^5$ &  $\omega^5$ &  $0$ &  $0$ &  $0$ &  $0$ &  $0$ &  $0$ &  $0$ \\
21 &  $0$ &  $0$ &  $0$ &  $0$ &  $0$ &  $0$ &  $1$ &  $0$ &  $0$ &  $0$ &  $0$ \\
22 &  $0$ &  $0$ &  $0$ &  $0$ &  $0$ &  $0$ &  $0$ &  $1$ &  $0$ &  $0$ &  $0$ \\
23 &  $0$ &  $0$ &  $0$ &  $0$ &  $0$ &  $0$ &  $0$ &  $0$ &  $1$ &  $0$ &  $0$ \\
24 &  $0$ &  $0$ &  $0$ &  $0$ &  $0$ &  $0$ &  $0$ &  $0$ &  $0$ &  $1$ &  $0$ \\
25 &  $0$ &  $0$ &  $0$ &  $0$ &  $0$ &  $0$ &  $0$ &  $0$ &  $0$ &  $0$ &  $1$ \\
26 &  $0$ &  $0$ &  $0$ &  $0$ &  $0$ &  $1$ &  $0$ &  $0$ &  $\omega^4$ &  $\omega^4$ &  $1$ \\
27 &  $0$ &  $0$ &  $0$ &  $0$ &  $0$ &  $1$ &  $0$ &  $1$ &  $0$ &  $\omega^2$ &  $\omega^2$ \\
28 &  $0$ &  $0$ &  $0$ &  $0$ &  $0$ &  $1$ &  $0$ &  $\omega^4$ &  $1$ &  $0$ &  $\omega^4$ \\
29 &  $0$ &  $0$ &  $0$ &  $0$ &  $0$ &  $1$ &  $0$ &  $\omega^2$ &  $\omega^2$ &  $1$ &  $0$ \\
30 &  $0$ &  $0$ &  $0$ &  $0$ &  $0$ &  $0$ &  $0$ &  $1$ &  $\omega^4$ &  $1$ &  $\omega^4$ \\
31 &  $0$ &  $0$ &  $0$ &  $0$ &  $0$ &  $1$ &  $\omega$ &  $0$ &  $0$ &  $1$ &  $\omega^4$ \\
32 &  $0$ &  $0$ &  $0$ &  $0$ &  $0$ &  $1$ &  $-1$ &  $0$ &  $\omega^2$ &  $0$ &  $\omega^2$ \\
33 &  $0$ &  $0$ &  $0$ &  $0$ &  $0$ &  $1$ &  $\omega^5$ &  $0$ &  $1$ &  $\omega^2$ &  $0$ \\
34 &  $0$ &  $0$ &  $0$ &  $0$ &  $0$ &  $0$ &  $1$ &  $0$ &  $-1$ &  $\omega$ &  $\omega$ \\
35 &  $0$ &  $0$ &  $0$ &  $0$ &  $0$ &  $1$ &  $\omega^5$ &  $\omega^2$ &  $0$ &  $0$ &  $1$ \\
36 &  $0$ &  $0$ &  $0$ &  $0$ &  $0$ &  $1$ &  $-1$ &  $\omega^4$ &  $0$ &  $\omega^4$ &  $0$ \\
37 &  $0$ &  $0$ &  $0$ &  $0$ &  $0$ &  $0$ &  $1$ &  $\omega^5$ &  $0$ &  $-1$ &  $\omega^5$ \\
38 &  $0$ &  $0$ &  $0$ &  $0$ &  $0$ &  $1$ &  $\omega$ &  $1$ &  $\omega^4$ &  $0$ &  $0$ \\
39 &  $0$ &  $0$ &  $0$ &  $0$ &  $0$ &  $0$ &  $1$ &  $\omega$ &  $\omega$ &  $0$ &  $-1$ \\
40 &  $0$ &  $0$ &  $0$ &  $0$ &  $0$ &  $0$ &  $1$ &  $-1$ &  $\omega^5$ &  $\omega^5$ &  $0$ \\
\end{tabular}}}
\qquad\qquad\qquad\qquad\qquad\qquad\qquad\qquad\qquad\qquad\qquad\qquad
\subfloat[][]{
\label{D9_39_13_Bases}
\scalebox{0.8}{
\begin{tabular}{ccccccccc}
1 & 2 & 3 & 4 & 5 & 6 & 22 & 23 & 24 \\
1 & 2 & 3 & 4 & 27 & 31 & 34 & 37 & 38 \\
1 & 2 & 3 & 5 & 28 & 32 & 35 & 37 & 39 \\
1 & 2 & 3 & 6 & 29 & 33 & 36 & 38 & 39 \\
1 & 2 & 3 & 22 & 25 & 26 & 27 & 28 & 29 \\
1 & 2 & 3 & 23 & 25 & 30 & 31 & 32 & 33 \\
1 & 2 & 3 & 24 & 26 & 30 & 34 & 35 & 36 \\
1 & 7 & 8 & 9 & 10 & 11 & 22 & 23 & 24 \\
2 & 7 & 12 & 13 & 14 & 15 & 22 & 23 & 24 \\
3 & 8 & 12 & 16 & 17 & 18 & 22 & 23 & 24 \\
4 & 9 & 13 & 16 & 19 & 20 & 22 & 23 & 24 \\
5 & 10 & 14 & 17 & 19 & 21 & 22 & 23 & 24 \\
6 & 11 & 15 & 18 & 20 & 21 & 22 & 23 & 24 \\
\end{tabular}}}
\qquad\qquad\qquad\qquad\qquad
\subfloat[][]{
\label{D11_40_12_Bases}
\scalebox{0.8}{
\begin{tabular}{ccccccccccc}
1 & 2 & 3 & 4 & 5 & 25 & 29 & 33 & 36 & 38 & 40 \\
1 & 6 & 7 & 8 & 9 & 10 & 30 & 34 & 37 & 39 & 40 \\
2 & 6 & 11 & 12 & 13 & 14 & 30 & 34 & 37 & 39 & 40 \\
3 & 7 & 11 & 15 & 16 & 17 & 30 & 34 & 37 & 39 & 40 \\
4 & 8 & 12 & 15 & 18 & 19 & 30 & 34 & 37 & 39 & 40 \\
5 & 9 & 13 & 16 & 18 & 20 & 21 & 22 & 23 & 24 & 25 \\
5 & 9 & 13 & 16 & 18 & 20 & 30 & 34 & 37 & 39 & 40 \\
10 & 14 & 17 & 19 & 20 & 21 & 26 & 27 & 28 & 29 & 30 \\
10 & 14 & 17 & 19 & 20 & 22 & 26 & 31 & 32 & 33 & 34 \\
10 & 14 & 17 & 19 & 20 & 23 & 27 & 31 & 35 & 36 & 37 \\
10 & 14 & 17 & 19 & 20 & 24 & 28 & 32 & 35 & 38 & 39 \\
10 & 14 & 17 & 19 & 20 & 25 & 29 & 33 & 36 & 38 & 40 \\
\end{tabular}}}
\caption{\protect\subref{D9_39_13} \& \protect\subref{D9_39_13_Bases}: The 39 indexed projectors (shown as kets, with $\omega=e^{i \pi /3}$) and 13 contexts of the critical $39-13$ KS set in dimension $d=9$ obtained by applying the Matsuno method to the 6-dimensional $21-7$ KS set of Fig. 1b of the main text.  \protect\subref{D11_40_12} \& \protect\subref{D11_40_12_Bases}: The 40 indexed projectors and 12 contexts of the critical $40-12$ KS set in dimension $d=11$ obtained by a computational search of the $41-24$ given by applying the Matsuno method to the same 6-dimensional $21-7$ KS set.}
\end{figure}

\begin{figure}
\centering
\subfloat[][]{
\label{D3_49_36}
\scalebox{1}{
\begin{tabular}{c|ccccc|ccccc|ccc}
1 &  $1$ &  $0$ &  $0$ & $ $ $ $ & 18 &  $2$ &  $1$ &  $1$ & $ $ $ $ & 35 &  $1$ &  $\bar{5}$ &  $2$ \\
2 &  $0$ &  $1$ &  $0$ &  & 19 &  $2$ &  $1$ &  $\bar{1}$ &  & 36 &  $5$ &  $1$ &  $\bar{2}$ \\
3 &  $0$ &  $0$ &  $1$ &  & 20 &  $2$ &  $\bar{1}$ &  $1$ &  & 37 &  $1$ &  $5$ &  $2$ \\
4 &  $1$ &  $1$ &  $2$ &  & 21 &  $2$ &  $\bar{1}$ &  $\bar{1}$ &  & 38 &  $5$ &  $1$ &  $2$ \\
5 &  $1$ &  $\bar{1}$ &  $2$ &  & 22 &  $2$ &  $0$ &  $1$ &  & 39 &  $1$ &  $5$ &  $\bar{2}$ \\
6 &  $1$ &  $\bar{1}$ &  $\bar{2}$ &  & 23 &  $2$ &  $0$ &  $\bar{1}$ &  & 40 &  $5$ &  $\bar{1}$ &  $2$ \\
7 &  $1$ &  $1$ &  $\bar{2}$ &  & 24 &  $1$ &  $1$ &  $1$ &  & 41 &  $1$ &  $\bar{5}$ &  $\bar{2}$ \\
8 &  $1$ &  $0$ &  $2$ &  & 25 &  $1$ &  $1$ &  $\bar{1}$ &  & 42 &  $2$ &  $\bar{5}$ &  $\bar{1}$ \\
9 &  $1$ &  $0$ &  $\bar{2}$ &  & 26 &  $1$ &  $\bar{1}$ &  $1$ &  & 43 &  $2$ &  $5$ &  $\bar{1}$ \\
10 &  $0$ &  $1$ &  $2$ &  & 27 &  $1$ &  $\bar{1}$ &  $\bar{1}$ &  & 44 &  $2$ &  $\bar{5}$ &  $1$ \\
11 &  $0$ &  $1$ &  $\bar{2}$ &  & 28 &  $1$ &  $1$ &  $0$ &  & 45 &  $2$ &  $5$ &  $1$ \\
12 &  $1$ &  $2$ &  $1$ &  & 29 &  $1$ &  $\bar{1}$ &  $0$ &  & 46 &  $5$ &  $\bar{2}$ &  $1$ \\
13 &  $1$ &  $2$ &  $\bar{1}$ &  & 30 &  $1$ &  $0$ &  $1$ &  & 47 &  $5$ &  $2$ &  $\bar{1}$ \\
14 &  $1$ &  $\bar{2}$ &  $\bar{1}$ &  & 31 &  $1$ &  $0$ &  $\bar{1}$ &  & 48 &  $5$ &  $\bar{2}$ &  $\bar{1}$ \\
15 &  $1$ &  $\bar{2}$ &  $1$ &  & 32 &  $0$ &  $1$ &  $1$ &  & 49 &  $5$ &  $2$ &  $1$ \\
16 &  $0$ &  $2$ &  $1$ &  & 33 &  $0$ &  $1$ &  $\bar{1}$ &  & & & & \\
17 &  $0$ &  $2$ &  $\bar{1}$ &  & 34 &  $5$ &  $\bar{1}$ &  $\bar{2}$ &  &  & & & \\
\end{tabular}}}
\qquad
\subfloat[][]{
\label{D3_49_36_Bases}
\scalebox{1}{
\begin{tabular}{ccc|ccc|ccc|ccc}
1 & 2 & 3 & 4 & 23 & 35 & 7 & 22 & 41 & 11 & 14 & 49 \\
1 & 10 & 17 & 4 & 25 & 29 & 7 & 24 & 29 & 12 & 26 & 31 \\
1 & 11 & 16 & 5 & 16 & 36 & 8 & 19 & 42 & 13 & 27 & 30 \\
1 & 32 & 33 & 5 & 23 & 37 & 8 & 21 & 43 & 14 & 25 & 30 \\
2 & 8 & 23 & 5 & 27 & 28 & 9 & 18 & 44 & 15 & 24 & 31 \\
2 & 9 & 22 & 6 & 17 & 38 & 9 & 20 & 45 & 18 & 27 & 33 \\
2 & 30 & 31 & 6 & 22 & 39 & 10 & 13 & 46 & 19 & 26 & 32 \\
3 & 28 & 29 & 6 & 26 & 28 & 10 & 15 & 47 & 20 & 25 & 32 \\
4 & 17 & 34 & 7 & 16 & 40 & 11 & 12 & 48 & 21 & 24 & 33 \\
\end{tabular}}}
\qquad
\subfloat[][]{
\label{D3_57_40}
\scalebox{1}{
\begin{tabular}{c|ccccc|ccccc|ccc}
1 &  $1$ &  $0$ &  $0$ & $ $ $ $ $ $ & 20 &  $0$ &  $\sqrt{2}$ &  $1$ & $ $ $ $ $ $ & 39 &  $1$ &  $\bar{3}$ &  $\sqrt[-]{2}$ \\
2 &  $0$ &  $1$ &  $0$ &  & 21 &  $0$ &  $\sqrt{2}$ &  $\bar{1}$ &  & 40 &  $1$ &  $\bar{3}$ &  $\sqrt{2}$ \\
3 &  $0$ &  $0$ &  $1$ &  & 22 &  $1$ &  $\sqrt{2}$ &  $1$ &  & 41 &  $1$ &  $3$ &  $\sqrt{2}$ \\
4 &  $1$ &  $1$ &  $0$ &  & 23 &  $1$ &  $\sqrt{2}$ &  $\bar{1}$ &  & 42 &  $\sqrt{2}$ &  $1$ &  $\bar{3}$ \\
5 &  $1$ &  $\bar{1}$ &  $0$ &  & 24 &  $1$ &  $\sqrt[-]{2}$ &  $\bar{1}$ &  & 43 &  $\sqrt{2}$ &  $\bar{3}$ &  $1$ \\
6 &  $1$ &  $0$ &  $1$ &  & 25 &  $1$ &  $\sqrt[-]{2}$ &  $1$ &  & 44 &  $\sqrt{2}$ &  $1$ &  $3$ \\
7 &  $1$ &  $0$ &  $\bar{1}$ &  & 26 &  $1$ &  $0$ &  $\sqrt{2}$ &  & 45 &  $\sqrt{2}$ &  $\bar{3}$ &  $\bar{1}$ \\
8 &  $0$ &  $1$ &  $1$ &  & 27 &  $1$ &  $0$ &  $\sqrt[-]{2}$ &  & 46 &  $\sqrt{2}$ &  $\bar{1}$ &  $\bar{3}$ \\
9 &  $0$ &  $1$ &  $\bar{1}$ &  & 28 &  $0$ &  $1$ &  $\sqrt{2}$ &  & 47 &  $\sqrt{2}$ &  $3$ &  $1$ \\
10 &  $\sqrt{2}$ &  $1$ &  $0$ &  & 29 &  $0$ &  $1$ &  $\sqrt[-]{2}$ &  & 48 &  $\sqrt{2}$ &  $\bar{1}$ &  $3$ \\
11 &  $\sqrt{2}$ &  $\bar{1}$ &  $0$ &  & 30 &  $1$ &  $1$ &  $\sqrt{2}$ &  & 49 &  $\sqrt{2}$ &  $3$ &  $\bar{1}$ \\
12 &  $\sqrt{2}$ &  $0$ &  $1$ &  & 31 &  $1$ &  $\bar{1}$ &  $\sqrt{2}$ &  & 50 &  $3$ &  $1$ &  $\sqrt[-]{2}$ \\
13 &  $\sqrt{2}$ &  $0$ &  $\bar{1}$ &  & 32 &  $1$ &  $\bar{1}$ &  $\sqrt[-]{2}$ &  & 51 &  $3$ &  $\bar{1}$ &  $\sqrt{2}$ \\
14 &  $\sqrt{2}$ &  $1$ &  $1$ &  & 33 &  $1$ &  $1$ &  $\sqrt[-]{2}$ &  & 52 &  $3$ &  $\bar{1}$ &  $\sqrt[-]{2}$ \\
15 &  $\sqrt{2}$ &  $1$ &  $\bar{1}$ &  & 34 &  $1$ &  $\sqrt[-]{2}$ &  $3$ &  & 53 &  $3$ &  $1$ &  $\sqrt{2}$ \\
16 &  $\sqrt{2}$ &  $\bar{1}$ &  $1$ &  & 35 &  $1$ &  $\sqrt[-]{2}$ &  $\bar{3}$ &  & 54 &  $3$ &  $\sqrt[-]{2}$ &  $\bar{1}$ \\
17 &  $\sqrt{2}$ &  $\bar{1}$ &  $\bar{1}$ &  & 36 &  $1$ &  $\sqrt{2}$ &  $\bar{3}$ &  & 55 &  $3$ &  $\sqrt[-]{2}$ &  $1$ \\
18 &  $1$ &  $\sqrt{2}$ &  $0$ &  & 37 &  $1$ &  $\sqrt{2}$ &  $3$ &  & 56 &  $3$ &  $\sqrt{2}$ &  $1$ \\
19 &  $1$ &  $\sqrt[-]{2}$ &  $0$ &  & 38 &  $1$ &  $3$ &  $\sqrt[-]{2}$ &  & 57 &  $3$ &  $\sqrt{2}$ &  $\bar{1}$ \\
\end{tabular}}}
\qquad
\subfloat[][]{
\label{D3_57_40_Bases}
\scalebox{1}{
\begin{tabular}{ccc|ccc|ccc|ccc}
1 & 2 & 3 & 4 & 31 & 32 & 12 & 32 & 38 & 17 & 18 & 48 \\
1 & 8 & 9 & 5 & 30 & 33 & 12 & 33 & 39 & 17 & 26 & 49 \\
1 & 20 & 29 & 6 & 23 & 24 & 13 & 30 & 40 & 20 & 31 & 50 \\
1 & 21 & 28 & 7 & 22 & 25 & 13 & 31 & 41 & 20 & 33 & 51 \\
2 & 6 & 7 & 8 & 15 & 16 & 14 & 19 & 42 & 21 & 30 & 52 \\
2 & 12 & 27 & 9 & 14 & 17 & 14 & 27 & 43 & 21 & 32 & 53 \\
2 & 13 & 26 & 10 & 24 & 34 & 15 & 19 & 44 & 22 & 29 & 54 \\
3 & 4 & 5 & 10 & 25 & 35 & 15 & 26 & 45 & 23 & 28 & 55 \\
3 & 10 & 19 & 11 & 22 & 36 & 16 & 18 & 46 & 24 & 29 & 56 \\
3 & 11 & 18 & 11 & 23 & 37 & 16 & 27 & 47 & 25 & 28 & 57 \\
\end{tabular}}}
\caption{\protect\subref{D3_49_36} \& \protect\subref{D3_49_36_Bases}: The 49 indexed projectors (shown as kets, where an overbar indicates a negative sign) and 36 contexts of the smallest critical KS set known in dimension $d=3$.  Completing the contexts of the set of 31 projectors discovered by Kochen and Conway gives a noncritical $61-46$ set from which this critical set was obtained using a computational search.  \protect\subref{D3_57_40} \& \protect\subref{D3_57_40_Bases} The 57 indexed projectors and 40 contexts of the critical KS set in $d=3$ obtained by completing the contexts of the 33 projectors discovered by Peres.}
\end{figure}